\documentclass{ws-ijmpa-modified}
\usepackage[rflt]{floatflt}

\begin{document}

\markboth{Andreas Warburton (for the CDF Collaboration)}
{Branching Fractions and $CP$ Asymmetries
  in Two-Body Charmless $b$-Hadron Decays}

%
\catchline{}{}{}{}{}
%


\title{BRANCHING FRACTIONS AND $CP$ ASYMMETRIES IN TWO-BODY
  NONLEPTONIC CHARMLESS $b$-HADRON DECAYS}


\author{\footnotesize ANDREAS WARBURTON \\
(representing the CDF Collaboration)}


\address{Department of Physics, McGill University\\ 3600 rue University,
Montr\'eal, Qu\'ebec  H3A~2T8\\
Canada
}



\maketitle


\begin{abstract}
  Relative branching fractions of $B_{d,s}^0 \to h^+ \, h^{\prime -}$
  decays (where $h,h^\prime = K \,{\rm or}\,\pi$) and the direct $CP$
  asymmetry ${\cal A}_{CP}$ in the $B_d^0 \to K^+\,\pi^-$ mode are
  measured with $179\pm 11$~pb$^{-1}$ of data collected at $\sqrt{s} =
  1.96$~TeV using the CDF~II detector at the Fermilab Tevatron
  $p\bar{p}$ collider.  The first branching-fraction measurement of a
  $B_s^0$ meson to two pseudoscalars, ${\cal B}(B_s^0 \to K^+\,K^-)$,
  and a search for the baryon mode $\Lambda_b^0 \to p^+\,h^-$ are also
  presented, in addition to branching-fraction limits on the rare
  channels $B_s^0 \to K^+\,\pi^-$, $B_d^0\to K^+\,K^-$, and $B_s^0\to
  \pi^+\,\pi^-$.

\end{abstract}

\section{Introduction}

Two-body charmless hadronic decays of neutral $b$ hadrons can provide
insight into both the CKM matrix and possible new physics phenomena.
The CDF~II experiment\cite{cdf} can reconstruct significant samples of
these decay modes by virtue of the high yields of $b$-quark production
at the Fermilab Tevatron $p\bar{p}$ collider and the use of a
dedicated trigger on impact parameters of charged-particle tracks.

Whereas $B_d^0$ decays are also under intense scrutiny at the
$e^+\,e^- \to \Upsilon(4S)$ factories, studies of the $B_s^0$-meson
and $b$-baryon channels are presently unique to the Tevatron.
Properties of the heavier hadron states complement and extend
measurements in the $B_u^+$ and $B_d^0$-meson sectors, {\it e.g.}, by
constraining hadronic parameters that are difficult to
calculate\cite{theorists,neubert,buras}.

In this paper, I report new results on analyses of two-body charmless
hadronic decays of $B_d^0$, $B_s^0$, and $\Lambda_b^0$
hadrons\footnote{Except in the case of ${\cal A}_{CP}$,
  charge-conjugate modes are implied throughout.} in a sample with
time-integrated luminosity $179\pm 11$~pb$^{-1}$ collected by the
CDF~II detector between February 2002 and September 2003 at a
center-of-mass energy of $\sqrt{s} = 1.96$~TeV.

\section{Event Selection}

Data are collected using CDF's hadronic $B$ triggers, which are
designed to select events containing track pairs originating from a
common displaced vertex.  At the lowest trigger levels, pairs of
oppositely charged tracks are required to have each transverse momenta
$p_T \geq 2.0$~GeV/$c$ and total $p_T \geq 5.5$~GeV/$c$, impact
parameters $>$150~$\mu$m, and an azimuthal opening angle
between $20^\circ$ and $135^\circ$.  Candidate $b$ hadrons are then
reconstructed from the track pairs and required to have a transverse
decay length greater than 300~$\mu$m, their total momenta point back
to within 80~$\mu$m of the primary vertex, and an invariant mass
between 4.0 and 6.0~GeV/$c^2$.

\begin{floatingfigure}[r]{60mm}
\centerline{\psfig{file=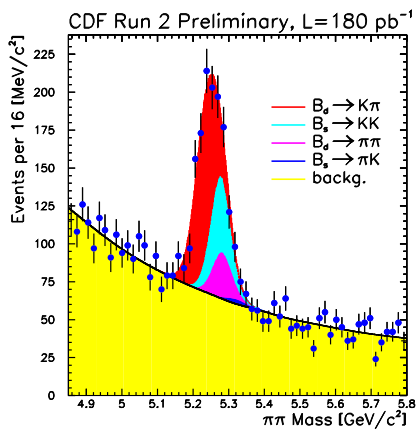,width=6cm}}
\vspace*{8pt}
\caption{The $M_{\pi\pi}$ invariant mass distribution of the $B^0\to
  h^+\, h^{\prime -}$ candidates after all selection criteria have
  been applied.  Overlaid are the fit results in this variable for the
  individual signal modes and the background component.}
\label{thefigure}
\end{floatingfigure}

Additional optimized cuts are applied offline to confirm the trigger
selection.  An isolation criterion, defined as $\frac{p_T(b)} {p_T(b)
  + \sum_i p_T({\rm track}_i)} \geq 0.5$, where the sum is over all
the non-$b$ tracks $i$ within a cone of radius\,=\,1 in
pseudorapidity-azimuth space around the candidate $b$-hadron
direction, is imposed to effect a $4\times$ background rejection
factor while retaining a signal efficiency of $\sim$85\%.  Specific
ionization ($dE/dx$) information measured in CDF's drift chamber is
used statistically in the fit to distinguish $K$ from $\pi$ mesons
with a separation of $1.39\sigma$ and $1.43\sigma$ for negative and
positive tracks, respectively.  This resolution suffices to provide a
separation between $K$ and $\pi$ mesons that is $\sim$60\% of what a
perfect particle identification scheme would achieve.

The di-track invariant mass distribution, where the world-average
$\pi$-meson mass\cite{pdg} is assigned to each daughter regardless of
$dE/dx$ information, is depicted by the points in
Fig.~\ref{thefigure}.  A clear signal of $893 \pm 47\,({\rm stat.})$
events is observed.

\section{Separation of Signal Modes and Background}

An unbinned maximum likelihood fit that combines `dipion' mass
$M_{\pi\pi}$, additional kinematic (kin), and $dE/dx$ particle
identification (PID) information is used to discriminate between the
two $B_d^0$ modes, the two $B_s^0$ modes, and the background
component.  The likelihood function used is ${\cal L} =
\prod_{j=1}^{N}{\cal L}_j$, where the index $j$ runs over the $N$
events and ${\cal L}_j$ has contributions from signal (sig) and
background (bkg): ${\cal L}_j = b \cdot {\cal L}^{\rm bkg} + (1 - b)
\cdot {\cal L}^{\rm sig}$, where $b$ is the background fraction.

The signal likelihood function is given by ${\cal L}^{\rm sig} =
\sum_k \xi_k \cdot {\cal L}_k^{\rm kin} \cdot {\cal L}_k^{\rm PID}$,
where the index $k$ runs over the six distinct $B_{d,s}^0$ decay-mode
and charge-state combinations, and the parameters $\xi_k$ represent
their relative fractions determined by the fit.  The background
likelihood function is expressed as ${\cal L}^{\rm bkg} = {\cal
  L}_{\rm bkg}^{\rm kin} \cdot {\cal L}_{\rm bkg}^{\rm PID}$.

In addition to $M_{\pi\pi}$, the other kinematic variable used is the
charge-signed momentum imbalance, defined as $\alpha \equiv Q_1 \cdot
\left(1 - \frac{p_1}{p_2}\right)$, where $p_1$($p_2$) is the modulus of the
smaller(larger) momentum of the two tracks and $Q_1$ is the charge
sign of the track assigned to $p_1$.  Analytic expressions for the
$M_{\pi\pi}$ dependence on $\alpha$ exhibit discriminating shape
differences between particles and antiparticles; self tagging is
thereby facilitated in the fit.

\section{Summary of Results}

The branching-fraction ratio of the $B_d^0$ modes is measured to be
\begin{equation}
\frac{{\cal B}(B_d^0 \to \pi^+\,\pi^-)} {{\cal B}(B_d^0 \to K^+\,\pi^-)} =
0.24 \pm 0.06 \,({\rm stat.}) \pm 0.05 \,({\rm syst.}),
\end{equation}
which is in agreement with the current world average\cite{pdg}.

The direct $CP$ asymmetry, ${\cal A}_{CP}$, is measured to be
\begin{equation}
    \frac{N(\overline{B}^0_d\to K^-\pi^+) - N(B^0_d \to K^+\pi^-)}
       {N(\overline{B}^0_d\to K^-\pi^+) + N(B^0_d \to K^+\pi^-)}
  = -0.04 \pm 0.08\, ({\rm stat.})  \pm 0.01\, ({\rm syst.}),
\end{equation}
which is compatible with both zero and recent evidence for a finite
asymmetry reported by the BaBar and Belle
collaborations\cite{babelle}.

In the $B_s^0$ sector, two ratios of fragmentation fractions ($f_d$,
$f_s$) and branching fractions are measured:
\begin{equation}
  \frac{f_s \cdot {\cal B}(B_s^0 \to K^+\,K^-)} {f_d \cdot {\cal B}(B_d^0 \to
    K^+\,\pi^-)} =
  0.50 \pm 0.08 \,({\rm stat.}) \pm 0.09 \,({\rm syst.})
\label{eqn:bs1}
\end{equation}
\begin{equation}
  \frac{f_d \cdot {\cal B}(B_d^0 \to \pi^+\,\pi^-)} {f_s \cdot {\cal B}(B_s^0 \to
    K^+\,K^-)} =
  0.48 \pm 0.12 \,({\rm stat.}) \pm 0.07 \,({\rm syst.}),
\label{eqn:bs2}
\end{equation}
where in both cases the following assumptions are made: $\Gamma_s =
\Gamma_d$, the $CP$ content of $B_s^0 \to K^+\,K^-$ is 100\% dominated
by the $CP$-even short-lived component, and\cite{sm} $\Delta\Gamma_s /
\Gamma_s = 0.12\pm 0.06$.  The quantity $\Delta\Gamma_s / \Gamma_s$ is
assumed because this $CP$ $B_s^0$ eigenstate may not possess a
lifetime identical to the average $B_s^0$ lifetime measured using
other modes, including semileptonic decays.  The uncertainty on
$\Delta\Gamma_s / \Gamma_s$ is included in the systematics in
Eqs.~\ref{eqn:bs1} and \ref{eqn:bs2}.  Combining Eqs.~\ref{eqn:bs1}
and \ref{eqn:bs2} with world-average results\cite{pdg} for the light
$B$ mesons provides a first measure of the branching fraction
\begin{equation}
  {\cal B}(B_s^0 \to K^+\,K^-) = \left[ 34.3 \pm 5.5\,({\rm
    stat.}) \pm 5.2\,({\rm syst.}) \right] \times 10^{-6},
\end{equation}
which is almost twice that for the $B_d^0\to K^+\,\pi^-$
mode\cite{pdg}.  The branching-fraction comparison between these two
modes, which differ only by flavor of spectator quark, is compatible
with expectations from QCD sum-rule calculations\cite{buras,khodj}.

The other $B_s^0$ channel for which an accessible branching fraction is
expected, $B_s^0 \to K^+\pi^-$, is not observed and a limit is therefore
set:
\begin{equation}
  \frac{f_s \cdot {\cal B}(B_s^0 \to K^+\,\pi^-)} {f_d \cdot {\cal B}(B_d^0 \to
    K^+\,\pi^-)} < 0.11 \; (90\%\; CL),
\end{equation}
which translates\cite{pdg} to
${\cal B}(B_s^0 \to K^+\,\pi^-) < 7.5 \times 10^{-6} \; (90\%\; CL)$,
near the lower end of current expectations\cite{neubert}.

Other rare modes, thought to be dominated by annihilation and exchange
processes and exhibiting branching fractions in an expected
range\cite{neubert,li} of $10^{-8}$ to $10^{-7}$, were sought by
adding contribution components for them to the likelihood expression.
Since the fit parameters changed negligibly after their inclusion, no
evidence for these modes was established, and the following limits
were set:
\begin{equation}
  \frac{{\cal B}(B_d^0 \to K^+\,K^-)} {{\cal B}(B_d^0 \to
    K^+\,\pi^-)} < 0.17 \; (90\%\; CL),
\end{equation}
or\cite{pdg}
${\cal B}(B_d^0 \to K^+\,K^-) < 3.1 \times 10^{-6} \; (90\%\; CL)$.
The current best limit measurement\cite{pdg} for this mode is $0.6
\times 10^{-6}$.  Also,
\begin{equation}
  \frac{{\cal B}(B_s^0 \to \pi^+\,\pi^-)} {{\cal B}(B_s^0 \to
    K^+\,K^-)} < 0.10 \; (90\%\; CL),
\end{equation}
which takes the assumption that both modes have the same average
lifetime and translates to ${\cal B}(B_s^0 \to \pi^+\,\pi^-) < 3.4
\times 10^{-6} \; (90\%\; CL)$.  This represents a significant
improvement over the previous best limit measurement\cite{pdg} of $170
\times 10^{-6}$.

Our data sample was also used separately to search for the charmless
two-body baryon decays\cite{mohanta} $\Lambda_b^0 \to p\,\pi^-$ and
$\Lambda_b^0 \to p\,K^-$ by counting the number of candidates
populating a search region with invariant mass between\cite{punzi}
5.415 and 5.535~GeV/$c^2$.  No evidence for these modes was found, and
an upper limit is therefore presented:
\begin{equation}
  \left[ {\cal B}(\Lambda_b^0 \to p\,\pi^-) +
         {\cal B}(\Lambda_b^0 \to p\, K^-) \right]
   < 22 \times 10^{-6} \; (90\%\; CL),
\end{equation}
which improves upon the previous limit\cite{pdg} of ${\cal B}(\Lambda_b^0
\to p\,h^-) < 50 \times 10^{-6}$.

\end{document}